\documentclass[twocolumn,aps,prc,superscriptaddress,showpacs]{revtex4}
\usepackage{graphicx}
\usepackage{epstopdf}
\usepackage{pdfpages}
\usepackage{subfigure}
\usepackage{CJK}

\usepackage{colordvi}
\begin{document}

\title{Collective flows of $^{16}$O + $^{16}$O collisions with $\alpha$-clustering configurations}
\author {Chen-Chen Guo}
\affiliation{ Shanghai Institute of Applied Physics, Chinese Academy of Sciences, Shanghai 201800, China}
\author {Wan-Bing He}
\affiliation{Institute of Modern Physics, Fudan University, Shanghai 200433, China}
\author {Yu-Gang Ma} \thanks{Email: Corresponding author.  ygma@sinap.ac.cn}
\affiliation{ Shanghai Institute of Applied Physics, Chinese Academy of Sciences, Shanghai 201800, China}
\affiliation{ University of Chinese Academy of Sciences, Beijing 100049, China }
\affiliation{ ShanghaiTech University, Shanghai 200031, China}
\date{\today}

\begin{abstract}
The main purpose of the present paper is to discuss whether or not the collective flows in heavy-ion collision at Fermi energy can be taken as a tool to investigate the cluster configuration in light nuclei. In practice, within an Extended Quantum Molecular Dynamics model, four $\alpha$-clustering (linear chain, kite, square, and tetrahedron) configurations of
$^{16}$O are employed in the initialization,
$^{16}$O+$^{16}$O around Fermi energy (40 - 60 MeV$/$nucleon) with impact parameter 1 - 3 fm are simulated, the directed and elliptic flows are analyzed. It is found that
collective flows are influenced by the different $\alpha$-clustering configurations, and the directed flow of free protons is more sensitive to the initial cluster configuration than the elliptic flow. Nuclear reaction at Fermi energy can be taken a useful way to study cluster configuration in light nuclei.
\end{abstract}


\maketitle

With the rapid development in both theoretical and experimental methods, $\alpha$ clustering structure in light nuclei has attracted much more attention in recent decades~\cite{Review1,Review2,CaoXG}.
There are various theoretical models to study $\alpha$ cluster in light nuclei. For example, the \emph{ab~initio} method \cite{Epelbaum:2011md,Epelbaum:2012qn}, the Fermion Molecular Dynamics model (FMD)~\cite{Feldmeier:1989st,Chernykh:2007zz}, the Antisymmetric Molecular Dynamics model (AMD)~\cite{KanadaEn'yo:2012bj,Kanada,MaCW,MaCW2}, the extended Quantum Molecular Dynamics model (EQMD) \cite{He:2014iqa,He:2016cwt,Huang,SHAN:2015,SHAN:2015b}, the $\alpha$-cluster model~\cite{Halcrow:2016spb}, the algebraic cluster model \cite{Bijker:2010zz}, the Covariant density functional theory~\cite{Zhou:2015nza} and so on. Regarding the $\alpha$-clustering configuration in $\rm{^{16}O}$, the  chain configuration of 4$\alpha$ clusters in $\rm{^{16}O}$ was investigated using a Skyrme cranked Hartree-Fock method~\cite{Ichikawa:2011iz} as well as the Brink wave functions \cite{Suhara}, while a tetrahedral configuration of $\alpha$ clusters was discussed in Refs.~\cite{Epelbaum:2013paa,Bijker:2014tka}.
In the experimental point of view, however, the whole picture
of $\alpha$-clustering configuration in nucleus is still not emerged so far although some experimental signatures have been indicated for information of $\alpha$ clusters in nuclei \cite{Review2,Bijker:2014tka,PKU}.
It is necessary to find more observables to explore cluster configuration in nuclei.

Recently, it was proposed that using collective flows in relativistic nuclear collisions to probe $\alpha$-clustering in light nuclei, which offers a new idea for investigating the cluster configuration~\cite{Broniowski:2013dia,ZhangS,Bozek:2014cva}.
Collective flow which has been studied over a wide range of beam energies and reaction systems,
is one of the most important observable to probe the nuclear equation of state, the in-medium nucleon-nucleon cross section, the quark-gluon plasma, the viscosity and so on~\cite{reisdorf1997,npa,Ma_flow,Kumar,Ko_flow,Yan,Chen,Han,CPL,CPL2,ZhouCL,Song,Huo,WangTT}. Thus, it is very interesting to evaluate whether it can be used to probe cluster configuration in light nuclei at Fermi energies.

In this work, within an extended quantum molecular dynamics (EQMD) model, the influence of cluster configuration on
collective flows of protons produced in $\rm{^{16}O}+\rm{^{16}O}$
collisions at the Fermi energy is investigated. The EQMD model is based on the
Quantum molecular dynamics (QMD) model which is a N-body approach
with considering several improvements, the approaches can be used to simulate nuclear reaction and meson-induced reaction at both very low energies and relativistic energies \cite{Aichelin:1991xy}. In QMD-type model, each nucleon is represented by a Gaussian wave packet. In the initialization of projectile and target in QMD-type models, The centers of the Gaussian wave packet of nucleons are randomly chosen in coordinate space between 0 and the radius of projectile or target as well as  momentum space between 0 and the Fermi momentum at the local density with considering several constraints, such as a proper binding energy, density and momentum distributions. However, the initializied nuclei are not always at their ground state (energy-minimum state), thus some unexpected nucleons are emitted during collision process even at zero temperature.
In order to solve those problems,
an extended version of QMD model was developed by Maruyama {\it et al.}, named EQMD~\cite{Maruyama:1995dc}. In the EQMD model, unlike the standard QMD model, the width of Gaussian wave packet of nucleon is a complex and has the form, i.e.
${\nu _i} \equiv \frac{1}{{{\lambda _i}}} + i{\delta _i}$,
where ${{\lambda _i}}$ and ${\delta _i}$ are its real and imaginary parts which are also time-dependent, respectively. The Gaussian wave packet is
\begin{equation}
{\phi _{i}}\left( {{\mathbf{r};t}} \right){ = }{\left( {\frac{{{\nu _{i}}{ + }{\nu _{i}}^ * }}{{2\pi }}} \right)^{{3 \mathord{\left/{\vphantom {3 4}} \right.\kern-\nulldelimiterspace} 4}}}\exp \left[ {{ - }\frac{{{\nu _{i}}}}{2}{{\left( {{\mathbf{r}}{ - }{\mathbf{r}_{i}(t)}} \right)}^2}{ + }\frac{{i}}{\hbar }{\mathbf{r}} \cdot {\mathbf{p}_{i}(t)}} \right]\
\end{equation}
here $\mathbf{r}_{i}(t)$ and $\mathbf{p}_{i}(t)$ are the centers of wave packet of nucleon \emph{i} in the coordinate and momentum space, respectively. The total wave function of a N-body system is assumed as the direct product of the Gaussian wave packet, i.e. $\Psi { = }\prod\limits_{i} {{\phi _{i}}} \left( {{\mathbf{r};t}} \right)$.
The Hamiltonian of the system is given as
\begin{eqnarray}
\begin{array}{l}
H = \left\langle {\Psi \left| {\sum\limits_i { - \frac{{{\hbar ^2}}}{{2{m}}}\nabla _i^2 - {{\hat T}_{c.m.}} + {{\hat H}_{ int }}} } \right|\Psi } \right\rangle \\
  = \sum\limits_i {\left[ {\frac{{\mathbf{p}_i(t)^2}}{{2m}} + \frac{{3{\hbar ^2}\left( {1 + \lambda _i^2\delta _i^2} \right)}}{{4m{\lambda _i}}}} \right]}  - {T_{c.m.}} + {H_{int}},
\end{array}
\end{eqnarray}
where ${T_{c.m.}}$ denotes the spurious zero-point center-of mass kinetic energy caused by the zero-point oscillation of the center-of-mass wave function of the clusters. The subtraction of the spurious zero-point center-of mass kinetic energy is necessary for reproducing binding energy of nuclei and cannot be neglected when one studies nuclear structure and reaction at low energies, more details about ${T_{c.m.}}$ can be found in Ref.\cite{ono:2004}. ${H_{ int }}$ is the potential energy term, it contains several parts as follows,
\begin{eqnarray}
{H_{{\mathop{\rm int}} }} = {H_{\rm{Skyrme}}} + {H_{\rm{Coulomb}}} + {H_{{\rm{Symmetry}}}}{\rm{ + }}{H_{{\rm{Pauli}}}},
\end{eqnarray}
where $H_{\rm{Skyrme}}$, $H_{\rm{Coulomb}}$, $H_{\rm{Symmetry}}$ and $H_{\rm{Pauli}}$ are the Skyrme, Coulomb,  Symmetry and Pauli potential terms. More details can be found in Ref.~\cite{Maruyama:1995dc}. Recently, the EQMD model was also extended to treat photonuclear reactions \cite{HuangBS,Huang,Wu_NST}.

In our previous studies, different $\alpha$-clustering structures and their effects on dipole resonance and photonuclear reactions of $^{16}$O and  $^{12}$C have been investigated \cite{He:2014iqa,He:2016cwt,Huang}. Here to study the influence of initial $\alpha$-cluster configurations on the collective flow, four configurations, namely linear chain, kite, square, and tetrahedron of 4-$\alpha$ structure for $^{16}$O are employed at the initialization, and random orientations are used to get the initial state for  $^{16}$O+$^{16}$O.  More than 600,000 events for each configuration are simulated, in order to reduce statistical error for observables.

\begin{figure}
\centering
\includegraphics[width=9cm,height=4.5cm]{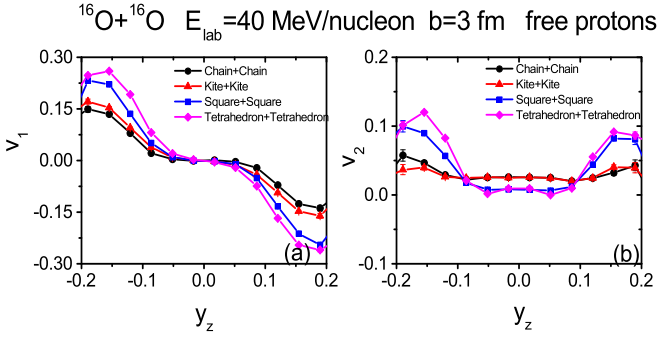}
\caption{\label{fig1}The directed flow parameter $v_1$ (a) and elliptic flow parameter $v_2$ (b) for free protons as a function of the longitudinal rapidity $y_z$ for
$^{16}$O + $^{16}$O collisions at 40 MeV$/$nucleon and  b = 3 fm.}
\end{figure}

\begin{figure}
\centering
\includegraphics[width=9cm,height=11cm]{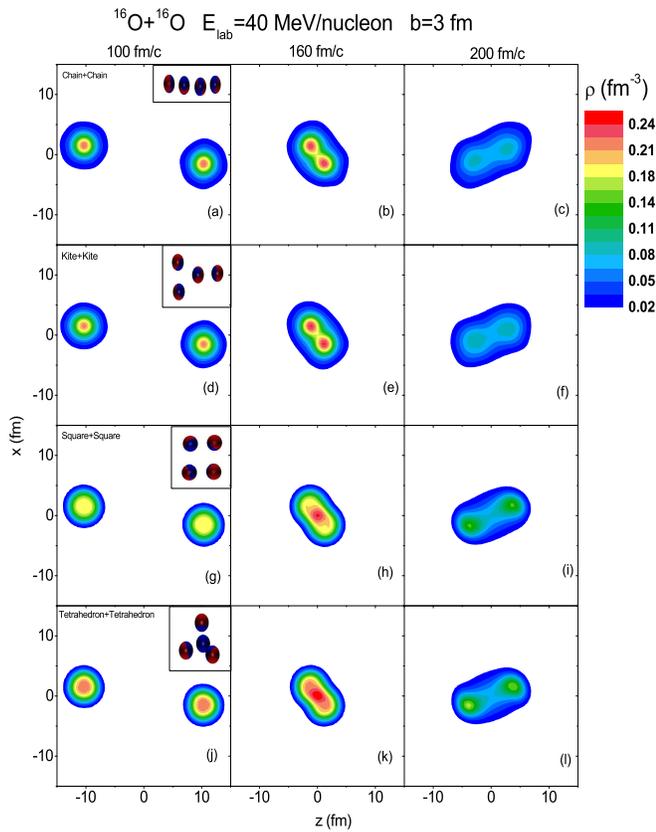}
\caption{\label{fig2} Contour plots of the average nucleon density in the reaction plane of $^{16}$O+$^{16}$O collisions at 40 MeV$/$nucleon. Simulations with the chain [panels (a), (b), and (c)], kite [panels (d), (e), and (f)], square [panels (g), (h), and (i)], and tetrahedron  [panels (j), (k), and (l)] configurations at three different times (100 fm$/$c, 160 fm$/$c, and 200 fm$/$c) are shown. In the inserts of (a), (d), (g) and (j),  sketches for 4 different $\alpha$-clustering configurations are plotted. 1000 events were collected to make the statistical error is small enough.
}
\end{figure}

Fig.\ref{fig1} shows the directed flow parameter $v_1 = \langle\frac{p_x}{p_t}\rangle$ and the elliptic flow parameter $v_2 = \langle\frac{p_x^2-p_y^2}{p_t^2}\rangle$ of free protons as a function of the longitudinal rapidity $y_z=\frac{1}{2}\ln\frac{E+p_z}{E-p_z}$. Here $p_t=\sqrt{p_x^2+p_y^2}$ is the
transverse momentum of emitted particles, $p_z$ is the $z$-component of momentum and $E$ is the total energy in the center-of-mass system. As usual, the \emph{x}-axis is defined to be along the impact parameter vector and the \emph{y}-axis perpendicular to that in the reaction plane, the \emph{z}-axis along the beam direction.) First, in Fig.\ref{fig1} (a) we clearly see that the directed flow parameter $v_1$ decreases with increasing rapidity (called negative flow, which means that particles are more likely to undergo a rotational-like motion rather than a bounce-off motion. The value of the elliptic flow parameter $v_2$ at mid-rapidity ($y_z$=0) are sightly larger than zero, as seen in Fig.\ref{fig1} (b), which implies a preferential in-plane emission rather than an out-of-plane emission pattern. Indeed, the observed negative directed flow and in-plane elliptic flow around the Fermi energy have been well established in both experimental and theoretical studies~\cite{reisdorf1997,Ma_flow,Andronic:2006ra,Li:2011zzp}, this results from the domination of attractive part of the nucleon-nucleon interaction over the repulsive part on the nucleon-nucleon scattering. Second, the $v_1$ of protons produced from the tetrahedron (more compact shape) configuration is much negative than that from the chain (less compact shape) configuration, because of a stronger interaction originated from a higher density in collisions with the former configuration (as will be seen in Fig.\ref{fig2}). In addition, we can see that, the differences in $v_1$ and $v_2$ between different cluster configurations become noticeable around the projectile/target rapidity region. From Fig.\ref{fig1} (b), we find that the calculated $v_2$ at mid-rapidity from the tetrahedron configuration is sightly smaller than that from the chain configuration, however, the situation is reverse around the target and projectile rapidities. This phenomenon may result from two factors: (i) relatively more violent two-body scatterings in the tetrahedron configuration determined by the initial geometry lead to smaller values of $v_2$ (towards out-of-plane emission) at mid-rapidity; (ii) weaker interactions exist in the quasi-projectile and quasi-target region in the chain configuration weaken the rotational-like motion of nucleons. In general, we note that the directed flow of free protons is more sensitive to the initial cluster configuration than the elliptic flow.

To better understand the influence of initial configuration on the reaction dynamic process, time evolution of the nucleon density in the reaction plane for $^{16}$O+$^{16}$O collisions at 40 MeV$/$nucleon is shown in Fig.\ref{fig2} as an example. At initial stage (here t=100 fm$/$c, because the projectile and target nuclei are put far away), see Figs.\ref{fig2}(a), (d), (g), and (j), the nucleon density contours for those four different configurations have almost spherical shapes, this is because the initial configuration are randomly orientated for each event, and 1000 events are accumulated to draw these contour plots. Nevertheless, the difference among them can be seen. It is understandable that, the contour of nucleon density for the chain configuration is the largest, while it for the tetrahedron configuration is the smallest. The difference in the density distribution at the initial stage will certainly affect the density evolution afterwards. As can be seen in the last two columns of Fig.\ref{fig2}, at t = 160 fm$/$c, the nucleon density in the center of the compressed region slightly increases from the chain to the tetrahedron configuration, and at t = 200 fm$/$c, the nucleon density in the center of the quasi-projectile and quasi-target region also increases from the chain to the tetrahedron configuration. Different densities achieved during the reaction would lead to different pressures, therefore, one can see that the final observables such as collective flows are influenced by the initial configuration.

\begin{figure}
\centering
\includegraphics[width=18cm,height=14cm]{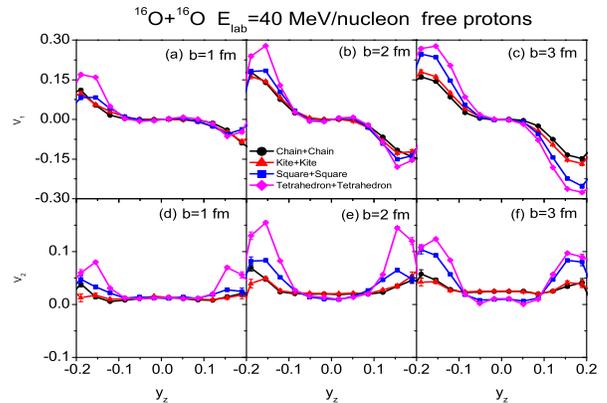}
\vspace{-9cm}
\caption{\label{fig3} The directed flow parameter $v_1$ (upper panels) and elliptic flow parameter $v_2$ (lower panels) of free protons in $^{16}$O+$^{16}$O collisions at 40 MeV$/$nucleon with impact parameter of 1, 2 and 3 fm, as a function of the longitudinal rapidity
$y_z$.}
\end{figure}

\begin{figure}
\centering
\includegraphics[width=18cm,height=14cm]{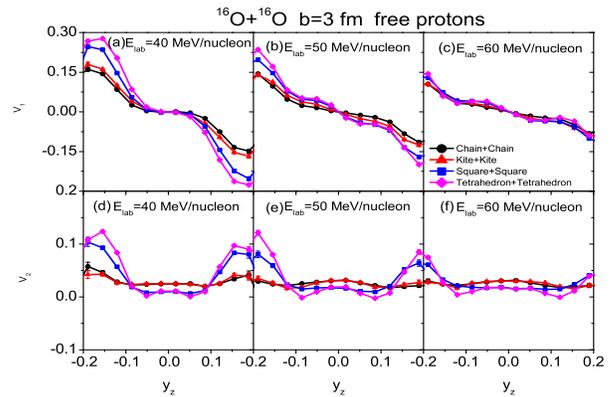}
\vspace{-9cm}
\caption{\label{fig4} Same as Fig.\ref{fig3}, but for $^{16}$O+$^{16}$O collisions at beam energies of 40, 50 and 60 MeV$/$nucleon with impact parameter of b = 3 fm.}
\end{figure}

To exhibit more systematically the influence of cluster configuration on the collective flows, $^{16}$O+$^{16}$O reaction with different impact parameters and beam energies are focused on. The results of the directed and elliptic flows of free protons are shown in Figs.\ref{fig3} and \ref{fig4}. As we can see from upper panels of Fig.\ref{fig3}, the difference between different cluster configurations is more evident for the larger impact parameters. We have checked that the difference in the nucleon density contour in the reaction plane increases with increasing impact parameter, which results in a larger sensitivity of the directed flow to the cluster configuration. For more central collision, the collision number is hardly influenced by cluster configuration due to random orientations of the initializied projectile and target, while for more peripheral collision, the collision number is strongly dependent on the initial geometry, thus the flow is sensitive to the cluster configuration.
The directed flow at different beam energies is shown in Fig.\ref{fig4}. The sensitivity of the directed flow to the cluster configuration decreases with increasing beam energies, e.g., at 60 MeV$/$nucleon, we can see almost the same directed flow with different cluster configurations. In both Fig.\ref{fig3} and Fig.\ref{fig4}, we find that the elliptic flow is less sensitive to different cluster
configurations than the directed flow. Beam energy below 40 MeV/nucleon is not considered, because nuclei are more likely to undergo fusion then the collective flows of nucleons become quite weak. The energy dependence of collective flow in HICs has been widely studied both experimentally and theoretically (see, e.g., \cite{Andronic:2006ra,Li:2011zzp}), with increasing beam energy, the multiple scattering process is dominant in comparison with the contribution coming from the effect of the clustering configuration.

In summary,  $^{16}$O+$^{16}$O reactions with  the linear chain, kite, square, and tetrahedron 4-$\alpha$ configurations around Fermi energy were simulated by the EQMD model and their  directed and elliptic flows of free protons are focused. It is found that
the directed flow of protons produced from the
 tetrahedron configuration is more negative than that from the
 chain configuration, while in-plan elliptic flow from the tetrahedron configuration is smaller than that from the chain configuration. This can be understood from the difference in the mean-field potential and two-body scattering caused by the initial geometry. Moreover, we also found that the directed flow of protons shows a larger sensitivity to the initial cluster configuration than the elliptic flow. In addition, through the simulation of $^{16}$O+$^{16}$O reaction at different impact parameters and beam energies, it is shown that the directed flow from reaction with a larger impact parameter and a smaller beam energy (around 40 MeV$/$nucleon) is suggested for future measurement to probe the cluster configuration in $^{16}$O.
From an experimental point of view, a precise measurement of the collective flow is relative easier for a larger colliding system due a better determination of the reaction plane. Thus, it would probably be advisable to investigate the cluster configuration with nuclear reaction such as, $^{16}$O ($^{12}$C) + $^{124}$Sn ($^{197}$Au). In a future study, such reactions will be simulated within the same microscopic transport model. Finally, we shall mention that here we did not touch the study of  high-order flows, such as triangular flow ($v_3$) and 4-th flow ($v_4$), which have been extensively studied in relativistic heavy ion collision \cite{Han,ZhangS}.  In principle, $v_3$ and  $v_4$ will be significant for nuclear reaction with triangle 3-$\alpha$ clustering nuclei ($^{12}$C) and 4-$\alpha$ clustering nucleus ($^{16}$O), respectively.  Anyway, such high-order flows leave us a topic to be addressed  in near future.

 This work was supported partly by the National Natural Science Foundation of China under Contract Nos. 11421505, 11220101005, 11305239, and 11605270, the Major State Basic Research Development Program in China under Contract  2014CB845401, and the Key Research Program of Frontier Sciences of the CAS under Grant No.
QYZDJ-SSW-SLH002, and by the China Postdoctoral Science Foundation Funded Project No. 2016M591730.
 We thank Prof. Qingfeng Li for helpful discussions and acknowledge support by the computing sever C3S2 in Huzhou University.

\end{document}